\documentclass[a4paper,11pt]{article}
\usepackage{pos}
\usepackage{placeins}

\title{The Future Circular Collider (FCC) at CERN}

\author*[a]{Rebeca Gonzalez Suarez}

\affiliation[a]{Department of Physics and Astronomy, Uppsala University\\
  Lägerhyddsvägen 1, 752 37 Uppsala}

\emailAdd{rebeca.gonzalez.suarez@physics.uu.se}

\abstract{With the LHC about to start its last data-taking period before being upgraded to the High-Luminosity LHC, it is time for the international high energy physics community to define the future of collider particle physics. The European Strategy for Particle Physics highlights an electron-positron Higgs boson factory as the main priority and as a first step towards a very high-energy future hadron collider. 

A staged Future Circular Collider (FCC), consisting of a luminosity-frontier highest-energy electron-positron collider (FCC-ee) followed by an energy-frontier hadron collider (FCC-hh), promises the most far-reaching physics program for the post-LHC era. FCC-ee is a precision instrument to study the Z, W, Higgs and top particles, and offers unprecedented sensitivity to signs of new physics. Most of the FCC-ee infrastructure can later be reused for the subsequent hadron collider, FCC-hh. 

The FCC-hh provides proton-proton collisions at a centre-of-mass energy of 100 TeV and can directly produce new particles with masses of up to several tens of TeV. This collider will also measure the Higgs self-coupling and explore the dynamics of electroweak symmetry breaking. Thermal dark matter candidates will be either discovered or conclusively ruled out by FCC-hh. 

Heavy-ion collisions and ep collisions (FCC-eh) further contribute to the breadth of the overall FCC program. The integrated FCC infrastructure will serve the particle physics community through the end of the 21st century. 

This presentation summarizes the feasibility of such a plan, possible implementation and conceptual designs of FCC-ee and FCC-hh, as well as physics potential. 

}

\FullConference{%
  7th Symposium on Prospects in the Physics of Discrete Symmetries (DISCRETE 2020-2021)\\
  29th November - 3rd December 2021\\
 Bergen, Norway}

\begin{document}
\maketitle

\section{Introduction}
The Standard Model (SM) of particle physics continues being in constant evolution hand in hand with available technology. It started back in the 1970s, with the discovery of the Neutral currents in Gargamelle (CERN) in 1973 and of charmed  particles (BNL, SLAC) in 1974-76. Since then a collection of particles has been discovered and we have greatly expanded our understanding of electroweak and strong interactions.  This evolution has only been possible due to the advances in technology and increasing accelerator energies.

The Large Hadron Collider (LHC) at CERN has been operative since 2009, and its last data-taking run before the planned high-luminosity upgrade, Run 3, is about to start in the coming weeks. This year we are also celebrating the 10th anniversary of the discovery of the Higgs boson. It is then time to take stock of the situation. Before Run 3, the general-purpose detectors at the LHC (ATLAS and CMS) have produced more than 1,000 papers each (and counting) LHCb is on 600. We found the Higgs boson: the first of its kind, a fundamental scalar particle, a neutral boson. We have performed a thorough testing of the SM without observing significant deviations from the SM. No other new particles have been found after the Higgs boson. 

The discovery of the Higgs boson at the LHC closes a central chapter of the SM, and poses new questions. Other questions have been open for a while, including but not limited to: 
\begin{itemize}
\item We can only describe three out of four forces, lacking a fundamental description of gravitation.
\item There are 3 generations of matter with an unexplained mass hierarchy that we cannot understand. 
\item Neutrinos have mass and we expected them to be massless.
\item We lack an explanation for the matter-antimatter unbalance of the Universe.
\item We do not understand the origin of dark matter.
\item We do not understand the origin of dark energy either. 
\end{itemize} 
Finally, the SM displays theoretical holes and discrepancies solved by fine tuning, and cannot be understood as complete. 

The answer to these questions will be found on energy- and intensity-frontier colliders. And now is time to plan the next facility. 

After the LHC its upgrade, the HL-LHC will take over. What will come after that we don't know yet. There are multiple options on the table, with different geometries (linear or circular) and different goals (precision or discovery) linked to different particles (leptons or hadrons), but at the end other factors such as location and funding will also play a role. For the first time in the history of the field however we have no clear energy scale to target. We have then no choice but to go for something versatile, broad, and as powerful as possible. And while we plan, we cannot forget the fact that we still have a new particle that could help chart future explorations, the Higgs boson. 

The 2020 Update of the European Strategy for Particle Physics~\cite{CERN-ESU-015} states that \textit{``An electron-positron Higgs factory is the highest-priority next collider. For the longer term, the European particle physics community has the ambition to operate a proton-proton collider at the highest achievable energy.''} and calls for \textit{``Europe, together with its international partners, should investigate the technical and financial feasibility of a future hadron collider at CERN with a centre-of-mass energy of at least 100 TeV and with an electron-positron Higgs and electroweak factory as a possible first stage. Such a feasibility study of the colliders and related infrastructure should be established as a global endeavor and be completed on the timescale of the next Strategy update..''}. This is the spirit that launched the Future Circular Collider Feasibility Study in the summer of 2021.

\section{The Future Circular Collider integrated program}
The Future Circular Collider (FCC) integrated program is a proposed post-LHC high-energy frontier circular colliders at CERN. It presents a comprehensive, cost-effective program maximizing physics opportunities inspired by the successful LEP-LHC (1976-2038?) program and providing a seamless continuation after it.

\begin{figure}[h!]
\centering
\includegraphics[width=0.8\textwidth]{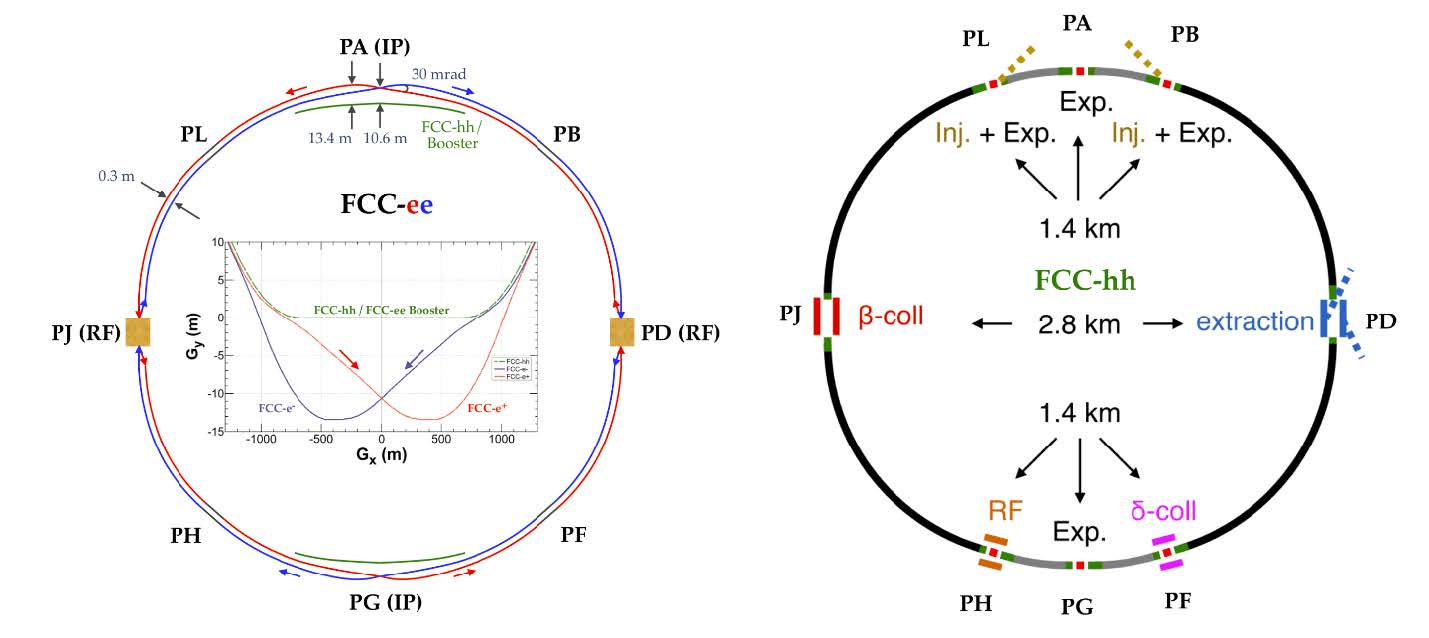}
\caption{The FCC-ee booster footprint coincides with that of the FCC-hh~\cite{FCC:2018evy}.}
\label{fig:foot}
\end{figure}

One tunnel of about 100~km of circumference will house the two stages of the project. Stage 1 will be an electron-positron collider, FCC-ee,   running with different center of mass energies to explore Z, W, H, and $t\bar{t}$ production. FCC-ee will be a Higgs, Electroweak, and top factory with high luminosity. Stage 2 will be a hadron collider, FCC-hh, with a center of mass energy of about 100~TeV. FCC-hh will be a natural continuation at energy frontier, with ion and eh options. These two phases will share infrastructure and will be implemented sequentially. They will benefit from common civil engineering and technical infrastructures and will build on and reuse CERN’s existing infrastructure as it is shown in Fig.~\ref{fig:foot}. 

A collider placement optimisation is underway following European and local regulatory frameworks and respecting a set of requirements and constraints, such as: civil engineering feasibility and subsurface constraints; territorial constraints at surface and subsurface nature, accessibility, technical infrastructure and resource needs and constraints; economic factors related to regional developments. The process is a collaborative effort by technical experts at CERN, consultancy companies and government notified bodies.

The current FCC layout baseline layout has 2 fallback solutions and proposes 8 pits (was 12) and a total circumference of 91.173 km. Stage 1, FCC-ee, could have 2 or 4 interaction points, and Stage 2, FCC-hh, is expected to have 4 interaction points. 

The FCC tunnel will have 5.5m inner diameter, which will be enough to fit the larger magnets of the hadron collider phase. Common experimental points are envisioned, with caverns large enough from the start to fit the larger hadron collider detectors (35~m high and wide). The proposed distance between detector cavern and service cavern is 50~m. 

The FCC feasibility Study started in 2021 and will go on for 5 years. If the project is approved before the end of this decade, construction can start at the beginning of the 2030s. FCC-ee operation could then span from 2045 to 2060, and FCC-hh operation from 2070 until 2090 or even further. 

The different collision types and center of mass energies expected in each different stage at the Future Circular Collider integrated program are presented in Table~\ref{tab:expect}.  

\begin{table}[h!]
\begin{center}
\caption{Stages, collision types, and center of mass energies expected for the Future Circular Collider integrated program.  \label{tab:expect}}
\begin{tabular}{ |c|c|c|c|c|} 
\hline
Stage & Collisions & $\sqrt{s}$ &  \\
\hline
\hline
FCC-ee & $e^+ e^-$ & 90~GeV (Z) & 2-4 IP \\ 
 & & 160~GeV (WW) & About 15 years of operation \\
 & & 240~GeV (H) & Very high luminosity Z pole run \\
 & & 3665 (tt) & (tera-Z) \\
 \hline
 \hline
FCC-hh & $pp$ & 100~TeV & 2+2 experiments\\ 
 & &  & 25 years of operation \\ 
  & $PbPb$ & 39~TeV & 1 run = 1 month of operation \\
\hline
FCC-eh & $ep$ & 3.5~TeV & Concurrent operation with pp \\
  & $ePb$ & 2.2~TeV & Concurrent operation with PbPb \\
\hline
 \end{tabular}
\end{center}
\end{table}

\FloatBarrier  

\subsection{FCC-ee}
\begin{figure}[h!]
\centering
\includegraphics[width=0.7\textwidth]{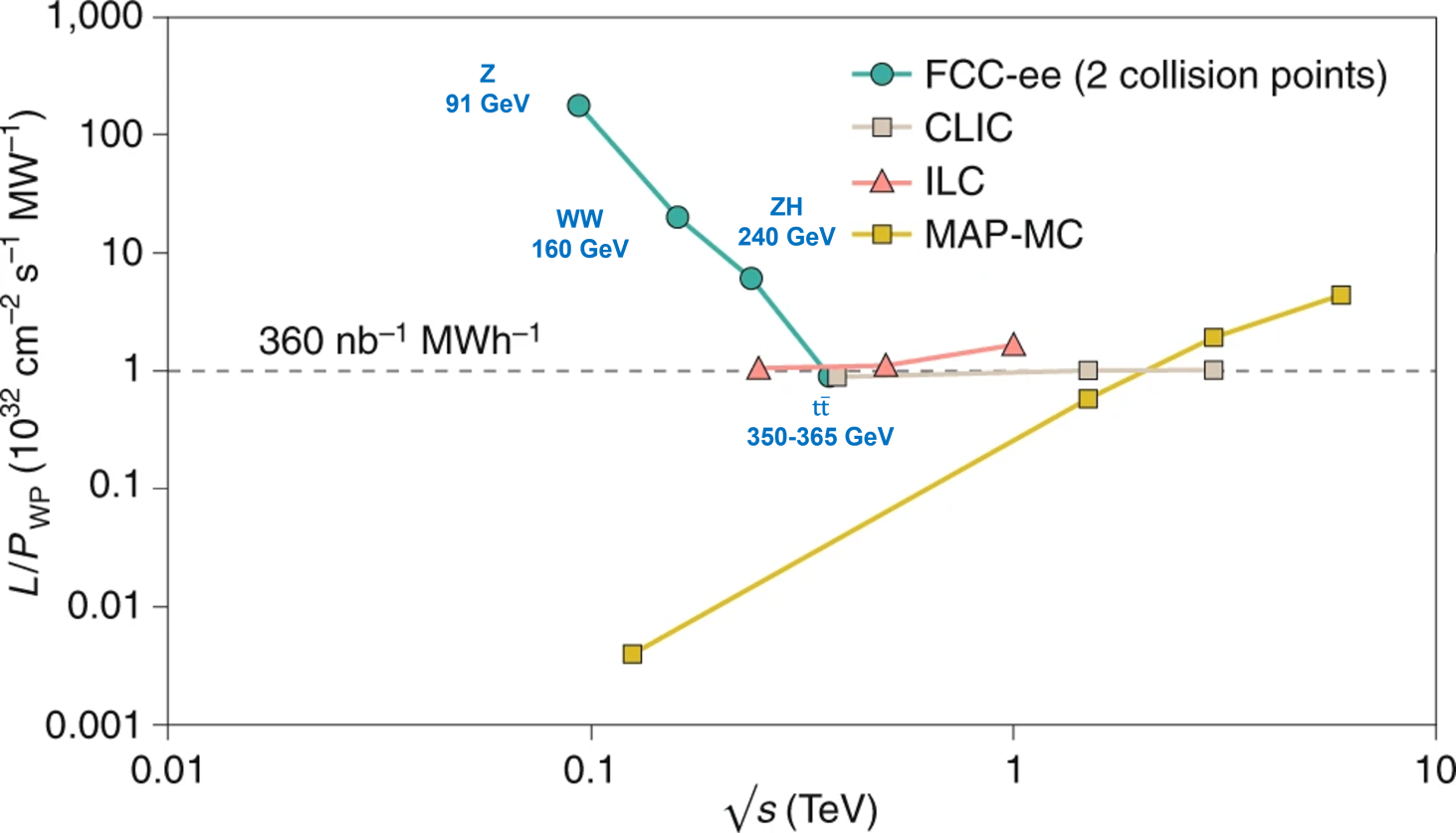}
\caption{Luminosity L per supplied electrical wall-plug power PWP is shown as a function of centre-of-mass energy for several proposed future lepton colliders. A value of 1 in the vertical units corresponds to 360$nb^{-1}$ MWh$^{-1}$ as indicated by the dashed line. Full caption and Figure from~\cite{naturefcc}.}
\label{fig:perf}
\end{figure}

The design of FCC-ee is based on lessons and techniques from past colliders. It displays a great energy range for the heavy particles of the SM. It offers complementarity with hadron (LHC, FCC-hh) and linear colliders. successful ingredients of several recent colliders are combined to achieve highest luminosities and energies. The expected luminosity and $\sqrt{s}$ for FCC-ee is presented in Fig.~\ref{fig:perf} compared with other proposed $e^+ e^-$ colliders.  

The FCC-ee will be implemented in stages as an electroweak, flavour, and Higgs factory to study with unprecedented precision the Higgs boson, the Z and W bosons, the top quark, and other particles of the Standard Model. More details are given in Table~\ref{tab:FCCee}.

\begin{table}[h!]
\begin{center}
\caption{FCC-ee phases.  \label{tab:FCCee}}
\begin{tabular}{ |c|c|c|c|c|c|} 
\hline
Phase & Run duration & $\sqrt{s}$ &  Integrated Luminosity & Event &  \\
 & (years) & (GeV) & ($ab^{-1}$) & stats. & \\
\hline
FCC-ee-Z & 4 & 88-95 & 150 & $3\times10^{12}$ visible Z decays & $10^5 \times$ LEP \\
FCC-ee-W & 2 & 158-162 & 12 & $10^8$ WW events &  $2\cdot 10^3 \times$ LEP \\
FCC-ee-H & 3 & 240 & 5 & $10^6$ ZH events &  Never done \\
FCC-ee-tt & 5 & 345-365 & 1.5 & $10^6$ $t\bar{t}$ events &  Never done \\
\hline
 \end{tabular}
\end{center}
\end{table}

\FloatBarrier

\subsection{FCC-hh}
\begin{figure}[h!]
\centering
\includegraphics[width=0.7\textwidth]{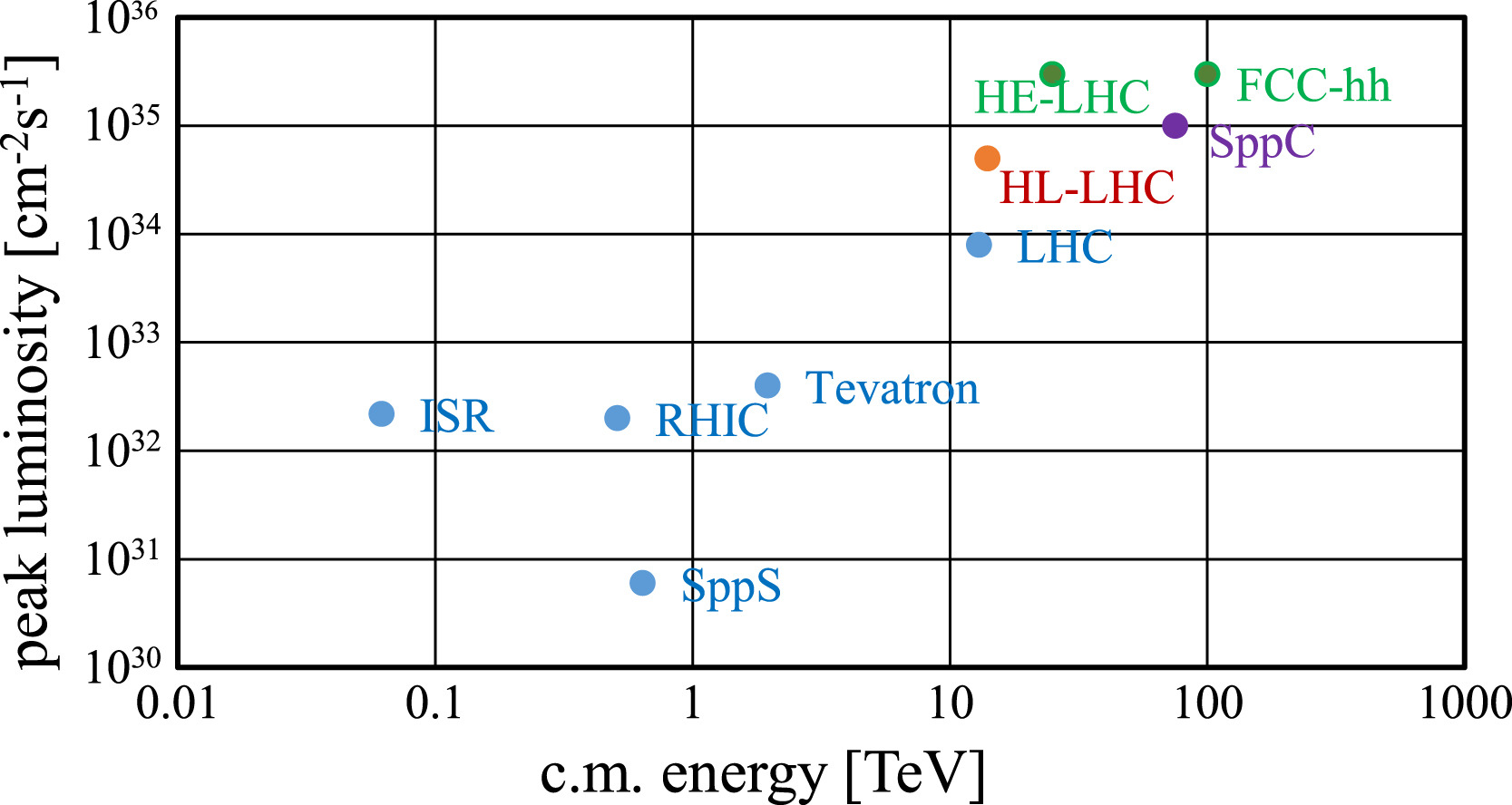}
\caption{
Luminosity vs. center-of-mass energy for past and present [blue], upcoming [red], and longer-term future hadron (or) colliders [green and purple] around the world~\cite{Benedikt:2018ofy}.}
\label{fig:perfhh}
\end{figure}

The highest collision energies will be achieved at FCC-hh. An order of magnitude performance increase in both energy and luminosity with respect to the LHC is expected, in a similar step as the one taken from the Tevatron to the LHC, with 100~TeV collision energy (vs. 14~TeV for LHC) and 20~$ab^{-1}$ per experiment over 25 years (vs 3~$ab^{-1}$ for LHC). 
This is presented in Fig.~\ref{fig:perfhh}. 

The key technology that will enable this step would be high-field magnets, from which there already exist a dipole demonstrator of 14.5~T made of NbSn at FNAL.

\FloatBarrier 

\subsection{Detector concepts}

There are currently two detector concepts for FCC-ee, that have been used for integration, performance, and cost estimates: one adapted from CLIC called CLD~\cite{Bacchetta:2019fmz}; and one specifically designed for FCC-ee (and CEPC) called IDEA~\cite{Antonello:2020tzq}.
 
Both are about 11 meters long and 6 meters high. CLD displays a full silicon tracker, 2T magnet field, high granularity ECAL (silicon-tungsten) and HCAL (scintillator-steel), and RPCs for muon detectors. IDEA on the other hand offers a silicon vertex detector, drift-chambers, and a dual-readout calorimeter (lead-scintillating, Cherenkov fibres). Complementary options are possible (especially with 4 experiments).

We are now taking a broader look at the physics potential and considering optimized detector designs for a complete physics program with many opportunities to contribute. 

\section{Physics}
A circular $e^+ e^-$ and $pp$ collider combined program will offer indirect high-mass-scale sensitivity together with direct search potential. It could provide the best possible precision and sensitivity for Higgs boson and top quark properties and electroweak symmetry breaking phenomena.

An integrated FCC program will have unprecedented direct and indirect exploration potential. 

The first stage, FCC-ee, is a Higgs factory, which means that the properties of the Higgs boson could be explored thoroughly.  Coupling deviations are likely to remain unconstrained at the HL-LHC, but that will not be the case after FCC-ee. The process $e^+e^- \rightarrow ZH$ will provide a model independent measurement of HZZ coupling. Measurement of the Higgs boson couplings to W, Z, b, $\tau$ will be possible below the \% level, as well as  \% to gluon and c.  It would be possible to perform absolute measurements of width and couplings thanks to the recoil method that allows to tag the Higgs boson as such independently of its decay mode.

The very large statistic expected for $pp \rightarrow H+X$ at FCC-hh together with the per-mille $e^+e^-$ measurement of Higgs properties available from FCC-ee, and large dynamic range will enable measurements of even rarer decay modes at less than 1\% level, like a $\leq 5\%$ measurement of the trilinear self-coupling. It will be possible to probe EFT operators of dimensions 5 and above up to scales of several TeV. Searches for multi-TeV resonances decaying to Higgs bosons, and extensions of the Higgs sector could be performed in a very large phase space. 

A circular $e^+e^-$ offers a clear advantage for the measurement of electroweak observables, a luminosity O($10^5$) with larger statistics than LEP at the Z peak and WW thresholds. Multiple properties will be measured to unprecedented precision: masses, asymmetries, branching ratios, or widths.

In terms of top and flavor physics, the top threshold region allows most precise measurements of top mass, width, and estimate of Yukawa coupling at FCC-ee. The study of this sector at FCC-hh has incredible potential but the reconstruction of these final states could be challenging.

The tera-Z run of FCC-ee will have 15 times the stats of Belle stats. This translates into great potential for studies of flavor anomalies. Large tau production is expected, in boosted topologies.  All b-hadron species will be available and there is clear potential for excellent secondary vertex reconstruction.

Experiments at FCC-ee can cover the full program of LHCb \& Belle II and compete favourably everywhere. 

Complementary Global EFT fits to EW and Higgs observables at FCC-ee will be possible.  Deviating operators may point to new physics to be found by the FCC-hh. 100~TeV is the appropriate energy to directly search new physics appearing indirectly through precision EW and Higgs measurements at FCC-ee.

Direct search at high scales will be the primary goal of FCC-hh, pushing the energy frontier directly.  All FCC stages however offer potential for direct searches, specially of new, feebly interacting particles that could manifest long-lived signatures. These signatures could be closely linked to dark matter, neutrino masses, or to the Baryon Asymmetry of the Universe (e.g. ALPs, exotic Higgs decays, or Heavy Neutral Leptons)~\cite{Chrzaszcz:2021nuk,Alimena:2022hfr}.

Heavy Neutral Leptons (HNL) are a good example in which to highlight the complementarity of the three FCC phases~\cite{Antusch:2016ejd}. FCC-ee will allow for indirect constrains from precision SM measurements and direct searches in single HNL production in Z decays. Sensitivity to $10^{-11}$ couplings for masses below the W mass. At FCC-hh, direct searches for single HNL production in W/Z decays will be possible together with studies of Lepton Number Violation and Lepton Flavor Violation. FCC-hh can test heavy neutrinos with masses up to about 2~TeV. Finally, FCC-eh can extend the reach of the FCC-hh up to about 2.7~TeV, providing the bestt reach above the W mass and sensitivity to LFV and Lepton-Number-violation signatures. 

Find more about the FCC in the European Strategy Update Documents~\cite{Benedikt:2653669, Benedikt:2653674,Benedikt:2653673}, the Conceptual Design Report~\cite{FCC-CDR2, FCCOp, FCC:2018vvp}, ~\cite{Blondel:2019ykp,Blondel:2019yqr,Blondel:2019qlh,Blondel:2019jmp}, and the latest snowmass report~\cite{Bernardi:2022hny}.

\section{Acknowledgements}
This talk was possible thanks to material from Michael Benedikt, Mogens Dam, Markus Klute, Fabiola Gianotti, Alain Blondel, Patrick Janot, Patrizia Azzi, Michelangelo Mangano, Gregorio Bernardi, and many others!

\bibliographystyle{abbrv}
\bibliography{references}

\end{document}